\documentclass[aps,prb,reprint,superscriptaddress,footinbib,citeautoscript]{revtex4-1}

\usepackage{graphicx}

\usepackage[american]{babel}
\usepackage[utf8]{inputenc}
\usepackage[T1]{fontenc}
\usepackage{enumerate}
\usepackage{mdwlist}
\usepackage[activate=normal]{pdfcprot}
\usepackage[babel]{csquotes}
\usepackage{bbding}
\usepackage{color}

\usepackage{amssymb}
\usepackage{amsmath}
\usepackage{amsfonts}
\usepackage{mathrsfs}
\usepackage{color}
\usepackage{graphicx}
\usepackage{bm}
\usepackage{dcolumn}

\usepackage{wrapfig}
\usepackage{color}

\newcommand* {\vek}[1]{{\ensuremath{\bm{\mathrm{#1}}}}}

\def\lsim{\raise0.3ex\hbox{$\;<$\kern-0.75em\raise-1.1ex\hbox{$\sim\;$}}}
\def\gsim{\raise0.3ex\hbox{$\;>$\kern-0.75em\raise-1.1ex\hbox{$\sim\;$}}}

\begin{document}
%

\title{RKKY interaction induced by two-dimensional hole gases}

\author{T. Kernreiter}
\affiliation{School of Chemical and Physical Sciences and MacDiarmid Institute for Advanced
Materials and Nanotechnology, Victoria University of Wellington, PO Box 600, Wellington
6140, New Zealand}

\begin{abstract}
We analytically compute the RKKY range function as induced by two-dimensional (2D) hole gases.
The bulk valence-band includes heavy-hole (HH) and light-hole (LH) states and their dynamics 
is described by the Luttinger Hamiltonian which we adopt as our framework. 
We show that even for situations where only the lowest HH-like subband is occupied   
the resulting form of the RKKY function can be very different as compared to the one of a 2D electron gas.
The associated spin susceptibility tensor has entries along the quantum-well directions
and perpendicular to it. Our formluae for the spin susceptibility tensor reveal 
the crucial influence of HH-LH mixing which gives rise to large anisotropies both among
the in-plane components as well as among the in-plane components and the component perpendicular to the quantum-well.
\end{abstract}

%
%
\maketitle
\section{Introduction}

The mechanism of indirect spin interaction of nuclei\cite{RK} or magnetic impurities\cite{K,Y} mediated by conduction electrons
has already been found in the 1950s and has been dubbed the Rudermann-Kittel-Kasuya-Yosida (RKKY) mechanism.
The corresponding effective Hamiltonian that describes the induced spin interaction of two magnetic impurities
is given by
\begin{equation}
\mathscr{H}_{\alpha\beta}^{\text{RKKY}}=-G^2 \sum_{i,j} S^{(\alpha)}_{i} S^{(\beta)}_{j} \chi_{ij}
({\vek{R}_\alpha,\vek{R}_\beta})~,
\label{eq:RKKYHam}
\end{equation}
where $S_{i}^{(\alpha)}$ denotes the $i$th Cartesian component of an impurity spin located at
position $\vek{R}_\alpha$, and $G$ is the exchange constant for the contact interaction
between the spin density of delocalized charge carriers with the impurity spins.
In Eq.~(\ref{eq:RKKYHam}), $\chi_{ij}({\vek{R}_\alpha,\vek{R}_\beta})$ is the spin
susceptibility which governs the form and range of the RKKY interaction and is determined
by quantities of the carrier system.
In the cases of an electron gas in three and two dimensions\cite{FischerKlein,Korenblit} the spin susceptibilities have a rather simple functional form with respect to the 
distance $R$ between two impurities, given as
\begin{equation}
\label{eq:RKKYfunc}
    \chi(R) \sim \begin{cases}
               \frac{1}{R^3}\left[\frac{\sin(2k_FR)}{2k_FR}-\cos(2k_FR)\right]             &(\text{3D})\\[3mm]
               J_0(k_FR)~Y_0(k_FR)+  J_1(k_FR)~Y_1(k_FR)     & (\text{2D})\\[2mm]
           \end{cases}
\end{equation}
Here $k_F$ is the Fermi wave vector and $J_n(\cdot)$ and $Y_n(\cdot)$ are Bessel functions of the first and second kind, respectively.
It follows from (\ref{eq:RKKYfunc}) that the associated Friedel oscillations
decay as $R^{-3}$ and $R^{-2}$ for a 3D and 2D electron gas (2DEG), respectively.

Ever since its discovery, the RKKY mechanism has been studied for a large variety of systems as 
it allows one not only to determine the spin orientation of two isolated impurities but still more importantly to obtain valuable 
information about the magnetic properties of a macroscopic system.
For the case of a 2DEG recent calculations have considered the influence of electron-electron interaction\cite{Simon},
Rashba\cite{Imamura,Huang,Lyu,Lai} spin-orbit coupling with Dresselhaus\cite{Mross,LossRKKY} spin-orbit coupling, and a
combination of electron-electron interaction and spin-orbit couplings\cite{Zak1,Zak2}.
For graphene it was noticed\cite{Brey1,Sherafati} that Friedel oscillations decay as $R^{-2}$ for the doped case and like $R^{-3}$ in the undoped case.

In the case of dilute magnetic semiconductors (DMS) it has been demonstrated \cite{Matsukura} that the experimentally found 
ferromagnetic order and transition temperatures can be ascribed to the RKKY mechanism. 
In this case, however, it is not mediated by conduction electrons but by valence holes. 
Subsequent theoretical studies\cite{Zarand,Priour2,Brey,Fiete,Timm} have considered various effects that can account for 
the observed magnitude of magnetization (for a recent review see Ref.~\onlinecite{Jungwirth}).
Also for two-dimensional DMS it has been suggested \cite{Haury,Dietl,Priour,Kechrakos,Melko,Meilikhov} that
the RKKY mechanism accounts for observed phenomena. 
The RKKY interaction in such two-dimensional hole systems is often modeled in the same fashion as in the case 
of electrons (\ref{eq:RKKYfunc}), assuming a one-band effective mass approximation.
Such assumptions, however, neglects the non-parabolic character\cite{Rashba} of hole dispersion bands and is therefore not always warranted.

Also taking into account the subtle effects due to non-parapolicity by means of a numerical subband k-dot-p theory\cite{Broido} 
calculation for a hole system based on GaAs, it has been shown\cite{KGZ} that the spin susceptibility tensor
exhibits strong anisotropy with the variation of the carrier density.
In particular, it has been pointed out that easy-plane entries of the spin susceptibility tensor can dominate over 
the easy-axis component. This feature was attributed to the effect of heavy hole (HH) light hole (LH) mixing 
which increases when the density of the hole gas is increased.

In the present paper, we provide further insight into the mechanism of HH-LH mixing and its influence on the RKKY range function,
where we give analytical results for the spin susceptibility tensor. This is advantageous as it allows us to retain
the explicit dependence on relevant band structure parameters. We base this calculation on an effective Luttinger model\cite{Pala,Bernevig,hole2DLind,Dollinger}, 
and demonstrate that the anisotropy of the spin susceptibility tensor entries
is intimately connected to the HH-LH mixed character of the hole states. 
Such an analytic result for the spin susceptibility tensor of two-dimensional hole gases is still missing in the literature
and deviates from the simple form of an equivalent electron system, Eq.~(\ref{eq:RKKYfunc}).

In Section \ref{sec:model}, we give a short account of the effective Luttinger model and define the relevant band structure parameters.  
In Section \ref{sec:spinsuscep}, we outline the calculation of the spin susceptibility tensor.
Numerical results are presented in Section \ref{sec:numerics}.
Section \ref{sec:sum} contains a short summary.

\section{Model\label{sec:model}}

In order to calculate the RKKY interaction mediated by 2D holes
our starting point will be the 4$\times$4 Luttinger model~\cite{Luttinger} as it 
provides a useful description of the upper-most valence band of typical semiconductors in situations where its couplings to the
conduction band and split-off valence band can be neglected. 
We adopt the Luttinger model in axial approximation, where we neglect anisotropic terms which are usually small:
\begin{subequations}
\begin{eqnarray}\label{eq:HL} 
&&\mathscr{H}_L =\mathscr{H}_0 + \mathscr{H}_1 + \mathscr{H}_2, \\[2mm]
&&\mathscr{H}_0 = -\frac{\hbar^2}{2m_0} \left[ \gamma_1 \left({\bf k}_\parallel^2 + k_z^2
\right) + \tilde\gamma_1 \left({\bf k}_\parallel^2 - 2 k_z^2 \right) \left(\hat J_z^2
-\frac{5}{4}\openone \right) \right], \nonumber \\ \\
&&\hskip-2mm\mathscr{H}_1 = \frac{\hbar^2}{m_0} \sqrt{2} \tilde\gamma_2 \left( \{ k_z , k_+ \}
\{\hat J_z , \hat J_- \} + \{ k_z , k_- \} \{\hat J_z , \hat J_+ \} \right), \nonumber \\ \\
&&\mathscr{H}_2 = \frac{\hbar^2}{2 m_0} \tilde\gamma_3 \left( k_+^2 \hat J_-^2 +
k_-^2 \hat J_+^2 \right).
\end{eqnarray}
\end{subequations}
Cartesian components of the spin-3/2 matrix vector are denoted by $\hat J_{x,y,z}$,
and we use the abbreviations $k_\pm = k_x \pm i k_y$, $\hat J_\pm = (\hat J_x
\pm i \hat J_y) / \sqrt{2}$, and $\{ A, B\} = (A B + B A)/2$. The constants
$\gamma_1$ and $\tilde\gamma_j$ are materials-dependent bandstructure
parameters~\cite{Vurgaftman}, where $\tilde\gamma_j$ depend also on the
quantum-well growth direction and their explicit expressions in terms of the standard
Luttinger parameters~\cite{Luttinger,Vurgaftman} $\gamma_2$ and $\gamma_3$
can be found, e.g., in Table~C.10 of Ref.~\onlinecite{RolandBook}.

A potential $V(z)$ along the $z$-direction models the confinement of holes to a 2D quantum well.
In the following, we assume the potential $V(z)$ to
be a hard-wall confinement with width $d$.
An effective Hamiltonian that describes the lowest size-quantized
orbital bound state approximately is then
obtained from (\ref{eq:HL}) by replacing $k_z \to \langle k_z \rangle = 0$ and $k_z^2
\to \langle k_z^2 \rangle = (\pi/d)^2$\cite{Pala,Bernevig,hole2DLind,Dollinger}. 
In such a way, we neglect HH-LH mixing among different orbital subbands.
In order to absorb the width dependence of our results into prefactors, we  
introduce the energy scale $E_0 =- \pi^2
\hbar^2 \gamma_1 / (2 m_0 d^2)$ and define wave vector components in units of $\pi/d$.
Throughout this paper we will work with dimensionless wave vectors and dimensionless
energies and include factors of $\pi/d$ and $E_0$ in the calculation where it is appropriate.
The (dimensionless) effective Hamiltonian is then given by
\begin{eqnarray}\label{eq:Luttinger2D}
&&\mathscr{H}^{{\rm 2D}}_L({\bf k}_\parallel)=E_0\left\{
\openone-2\bar{\gamma}\left(\hat{J}_z^2-\frac{5}{4}\openone\right)
\right.\nonumber\\
&&{}\left.
+\left[\openone+\bar{\gamma}\left(\hat{J}_z^2-\frac{5}{4}\openone\right)\right]{\bf k}_\parallel^2
-\alpha\bar{\gamma}\left(k_+^2 \hat{J}_-^2+k_-^2\hat{J}_+^2\right)
\right\}~,\nonumber\\
\end{eqnarray}
where we define the parameters $\bar\gamma \equiv \tilde\gamma_1/\gamma_1$ and
$\alpha \equiv \tilde\gamma_3/\tilde\gamma_1$ to discuss the effects of HH-LH splitting and HH-LH mixing
separately. Note that for ${\bf k}_\parallel=0$, the Hamiltonian (\ref{eq:Luttinger2D}) commutes with $\hat{J}_z$ which 
has eigenvalues $\pm 3/2$ (HH) and $\pm 1/2$ (LH).
Their corresponding energies are split up which is described by the parameter $\bar\gamma$.
For ${\bf k}_\parallel\neq 0$ (and $\alpha\neq0$), on the other hand, the eigenstates of the Hamiltonian 
are not simultaneously eigenstates of $\hat{J}_z$, with $\alpha$ describing the effect of HH-LH mixing.

\section{Spin susceptibility tensor\label{sec:spinsuscep}}

In the following we will calculate the spin susceptibility tensor for a 2D hole gas.
The analytical expression for the spin susceptibility tensor in linear response theory and
for finite temperature is conveniently given in terms of Matsubara Green's functions of the holes and reads\cite{Zyuzin}
\begin{eqnarray}\label{eq:Matsub}
\chi_{ij}({\bf R})=
k_{\text B}T\sum_n \text{Tr}\{\hat{J}_i G_{\omega_n}({\bf R}) \hat{J}_j G_{\omega_n}(-{\bf R})\}~,
\end{eqnarray}
where $\omega_n=(2n+1)\pi k_{\text B} T$ are the Matsubara frequencies. 
In the following we will consider the case of zero temperature for which
the result of the spin susceptibility tensor is straightforwardly obtained from Eq.~(\ref{eq:Matsub})
by making in the Green's functions the replacements $i\omega_n\to \omega +i\delta \text{sgn}(\omega)$ 
to obtain the retarded and advanced Green's functions for zero temperature\cite{Litvinov}.
Furthermore, we have to replace the sum by an integral according to $k_{\text B} T \sum_n\to \frac{1}{2\pi i}\int_{\Gamma_1}{\text d}\omega$,
with the contour of the integration given by $\Gamma_1=(-\infty-i\delta,-i\delta) \cup (i\delta,\infty+i\delta)$.
The Green's function in real space is calculated by a Fourier transformation of the Green's function
in momentum space, where the latter is given by
\begin{align}\label{eq:GreensfuncMom}
G_{\omega}({\bf k}_\parallel)=\frac{1}{E_0}\biggl[\bar{\omega}+\varepsilon_F-\mathscr{H}^{{\rm 2D}}_L({\bf k}_\parallel)/E_0\biggr]^{-1}~.
\end{align}
Here we use the abbreviation $\bar{\omega}\equiv\omega+i\delta {\rm sgn}(\omega)$, $\varepsilon_{F}$ is the Fermi energy,
and again we use dimensionless quantities as $\varepsilon_F\to E_0\varepsilon_F$ and $\bar\omega\to E_0\bar\omega$.

From Eq.~(\ref{eq:GreensfuncMom}) we obtain for the Green's function in momentum space (using polar coordinates):
\begin{eqnarray}\label{eq:GreensfuncMom1}
\left[G_{\omega}({\bf k}_\parallel)\right]_{ij}&=&
\left[A_-(\delta_{i1}+\delta_{i4})+A_+(\delta_{i2}+\delta_{i3})\right]\delta_{ij}\nonumber\\[2mm]
&&{}+B \left[{\rm e}^{-i2\phi_k}(\hat{J}_{+}^2)_{ij}+{\rm e}^{i2\phi_k}(\hat{J}_-^2)_{ij}\right]~,
\end{eqnarray}
with
\begin{eqnarray}
A_\mp&=&\frac{1}{E_0}\frac{1+k^2\mp\bar{\gamma}(k^2-2)-(\bar{\omega}+\varepsilon_{F})}{[\bar{\gamma}^2(1+3\alpha^2)-1](k^2-k_1^2)(k^2-k_2^2)}~,\qquad\nonumber\\
B&=&\frac{1}{E_0}\frac{\alpha\bar{\gamma} k^2}{[\bar{\gamma}^2(1+3\alpha^2)-1](k^2-k_1^2)(k^2-k_2^2)}~,
\end{eqnarray}
and $\delta_{ij}$ being the Kronecker symbol.
The Green's function has poles at 
%
\begin{eqnarray}\label{eq:Poles}
&&k_{1,2}=\frac{1}{\sqrt{1-\bar{\gamma}^2(1+3\alpha^2)}}\biggl[
\bar{\omega}+\varepsilon_{F}-1-2\bar{\gamma}^2\nonumber\\
&&{}\mp\sqrt{(\bar{\omega}+\varepsilon_{F}-3)^2
+3\alpha^2[(\bar{\omega}+\varepsilon_{F}-1)^2-4\bar{\gamma}^2]}
\biggr]^{1/2},
\end{eqnarray}
%
which coincide with the Fermi wave vectors\cite{hole2DLind} of the two hole states for $\bar\omega=0$.  
Thus, in order to obtain the Green's function in real space, we have to evaluate integrals of the form
\begin{eqnarray}
\{\mathscr{I},\mathscr{J},\mathscr{K}_\pm\}&=&\left(\frac{\pi}{d}\right)^2\frac{1}{E_0}\int_0^{2\pi}\frac{{\text d}\phi_k}{(2\pi)^2}
\int_0^{\infty}{\text d}k~k\nonumber\\[2mm]
&&{}\hskip-8mm\times\frac{\{1,k^2,k^2 {\text e}^{\pm i2\phi_k}\}~{\text e}^{i k R\cos(\phi_k-\phi_R)}}{(k^2-k^2_1)(k^2-k^2_2)}~,
\end{eqnarray}
where we use a dimensionless description also for the distance by changing $R\to (d/\pi) R$ and with
$\phi_R$ being the angle between the $x$-axis and the axis given by the two impurities. 
We calculate these integrals by using the Cauchy integral theorem, where we close the contour along the upper 
half-plane to obtain\cite{Dugaev}
\begin{eqnarray}\label{eq:Integrals}
\mathscr{I}&=&\frac{i}{4}\left(\frac{\pi}{d}\right)^2\frac{1}{E_0}
\frac{H^{(1)}_0(k_1R)-
H^{(1)}_0(k_2R)}{k_1^2-k_2^2}~,\nonumber\\[2mm]
\mathscr{J}&=&-\frac{i}{4}\left(\frac{\pi}{d}\right)^2\frac{1}{E_0}
\frac{k^2_1H^{(1)}_2(k_1R)-k^2_2H^{(1)}_2(k_2R)}{k_1^2-k_2^2}~,\nonumber\\[2mm]
\mathscr{K}_\pm&=&-\mathscr{J}{\rm e}^{\pm i2\phi_R}~.
\end{eqnarray}
Here $H^{(1)}_n(\cdot)$ denote Hankel functions of the first kind.
From Eqs.~(\ref{eq:GreensfuncMom1}) and (\ref{eq:Integrals}) we find for the Green's function in real space 
\begin{eqnarray}\label{eq:Geensfunreal}
\left[G_{\omega}({\bf R})\right]_{ij}&=&\left[\mathscr{A}_-(\delta_{i1}+\delta_{i4})+\mathscr{A}_+(\delta_{i2}+\delta_{i3})
\right]\delta_{ij}\nonumber\\[2mm]
&&{}+\mathscr{B}\left[{\rm e}^{-i2\phi_R}(\hat{J}_{+}^2)_{ij}+{\rm e}^{i2\phi_R}(\hat{J}_{-}^2)_{ij}\right],
\end{eqnarray}
with
\begin{eqnarray}
\mathscr{A}_\mp&=&\left(\frac{\pi}{d}\right)^2\frac{1}{E_0}\frac{[1\pm2\bar{\gamma}-(\bar\omega+\varepsilon_{F})]\mathscr{I}+(1\mp\bar{\gamma})\mathscr{J}}{\bar{\gamma}^2(1+3\alpha^2)-1}~,
\nonumber\\[2mm]
\mathscr{B}&=&-\left(\frac{\pi}{d}\right)^2\frac{1}{E_0}\frac{\alpha\bar{\gamma}\mathscr{J}}{\bar{\gamma}^2(1+3\alpha^2)-1}~,
\end{eqnarray}
where we have that $G_\omega(-{\bf R})=G_\omega({\bf R})$.
Performing the trace in Eq.~(\ref{eq:Matsub}) for the various non-vanishing entries of the susceptibility tensor, yields
\begin{subequations}
\begin{eqnarray}
\text{Tr}\{\hat{J}_{x} G_\omega \hat{J}_{x} G_\omega\}&=&3\mathscr{A}_
+\mathscr{A}_-+2 \mathscr{A}_+^2+9 \mathscr{B}^2
\nonumber\\
&&{}
+ 12\mathscr{A}_+\mathscr{B} \cos2\phi_R~,\label{eq:Trxx}\\[2mm]
\text{Tr}\{\hat{J}_{y} G_\omega \hat{J}_{y} G_\omega\}&=&3\mathscr{A}_
+\mathscr{A}_-+2 \mathscr{A}_+^2+9 \mathscr{B}^2
\nonumber\\
&&{}
- 12\mathscr{A}_+\mathscr{B} \cos2\phi_R~,\label{eq:Tryy}\\[2mm]
\text{Tr}\{\hat{J}_x G_\omega \hat{J}_y G_\omega\}&=&12\mathscr{A}_+
\mathscr{B} \sin2\phi_R~,\label{eq:Trxy}\\[2mm]
\text{Tr}\{\hat{J}_z G_\omega \hat{J}_z G_\omega\}&=&\frac{1}{2}\left[9\mathscr{A}_-^2
+\mathscr{A}_+^2-18 \mathscr{B}^2\right].\label{eq:Trzz}
\end{eqnarray}
\end{subequations}
Due to the appearance of the last term in Eqs.~(\ref{eq:Trxx}) and (\ref{eq:Tryy}) in-plane anisotropy of the RKKY interaction
is introduced by HH-LH mixing ($\alpha\neq 0$) which is a distinctive feature of a 2D hole system as compared to 
the corresponding electron system.

To obtain the final result for the spin susceptibility tensor, we still have to integrate the terms in Eqs.~(\ref{eq:Trxx})-(\ref{eq:Trzz}) over 
the frequency along the contour $\Gamma_1$.
For this integration, we use the method proposed in Ref.~\onlinecite{Litvinov}.
Within our framework, Eq.~(\ref{eq:Luttinger2D}), where we assume that only the lowest HH-like subband is occupied,
we can replace the integral along the contour $\Gamma_1$ by an integral 
along the two lines $\Gamma_2=(-i\delta,\omega_0-i\delta) \cup (\omega_0+i\delta,i\delta)$, with
$\omega_0=-\varepsilon_{F}+1-2\bar{\gamma}$.
Note that the two lines are below and above the branch cut in the complex frequency plane
which corresponds to the domain ($\omega_0$,$\infty$), where the real part under the square root in $k_2$ is positive.  
The possibility to exchange the integration domains is a consequence of Cauchy's integral theorem which states that the integral of an 
analytic function over a closed curve is zero, which means in our case
$\int_\cap+\int_{\Gamma_1}+\int_{\Gamma_2}=0$. The symbol $\cap$ denotes the curve in the upper half plane extended to infinity,
and the corresponding integral gives zero as the Hankel functions vanish in this limit.
Thus we can make the replacement $\int_{\Gamma_1}\to -\int_{\Gamma_2}$.  
We then evaluate all possible products of Hankel functions in Eqs.~(\ref{eq:Trxx})-(\ref{eq:Trzz}), where we find
\begin{subequations}
\begin{eqnarray}
&&\int_{\Gamma_2}{\text d}\omega f(\omega) H_n^{(1)}(k_1 R)H_m^{(1)}(k_2 R)=\nonumber\\[2mm]
&&{}i\frac{4}{\pi} {\text e}^{-i n \pi/2}\int^{\omega_0}_{0}{\text d} \omega  f(\omega)K_n(|k_1| R)J_m(k_2 R),\label{eq:IntOmega1} \quad \\[3mm] 
&&\int_{\Gamma_2}{\text d}\omega f(\omega) H_n^{(1)}(k_2 R)H_m^{(1)}(k_2 R)=\nonumber\\[2mm]
&&{}-2 i\int^{\omega_0}_{0}{\text d} \omega  f(\omega)\left[J_n(k_2 R)Y_m(k_j R)+J_m(k_2 R)Y_n(k_i R)\right],\nonumber\\ \label{eq:IntOmega2} 
\end{eqnarray}
\end{subequations}
for $n,m=0,2$ and $f(\omega)$ denotes an analytic function in $\omega$. 
In obtaining Eq.~(\ref{eq:IntOmega2}), we have used the relation 
$K_n(z)=\frac{i\pi}{2}{\text e}^{i n \pi/2} H_n^{(1)}(z {\text e}^{i\pi/2})$
for Hankel functions that have $k_1$ (which is imaginary) in their argument, where $K_n(\cdot)$ are the modified Bessel functions of the second kind.  
In addition we have used the relation
$H^{(1)}_n(z {\text e}^{i\pi})=-{\text e}^{-i\pi n}H^{(2)}_n(z)$ between Hankel functions of the first and second kind 
that contain $k_2$ and the definition of Hankel functions
in terms of Bessel functions. The integrated products of Hankel functions involving only $k_1$ give zero. 
Using Eqs.~(\ref{eq:Trxx})-(\ref{eq:Trzz}) together with Eqs.~(\ref{eq:IntOmega1}) and (\ref{eq:IntOmega2})
we finally obtain the (semi-)analytical result for the spin susceptibility tensor.


Considering the limit of large distances, $k_{F}R\gg1$, a particular simple expression can be found for the spin susceptibility
tensor, because in this case the Bessel functions can be approximated very well by 
\begin{eqnarray}
J_n(x)&\approx& \sqrt{\frac{2}{\pi x}}~\cos(x-n\pi/2-\pi/4)~,\nonumber\\
Y_n(x)&\approx& \sqrt{\frac{2}{\pi x}}~\sin(x-n\pi/2-\pi/4)~,
\end{eqnarray}
whereas $K_n(x)$ decays exponentially with the distance and can be approximated as $K_n(x)\approx 0$.
Using these approximations and setting $\phi_R=0$, the spin susceptibility tensor elements can be given by the compact expression
\begin{align}
\chi_{ii}({\bf R})=\chi_0\int_0^{\omega_0}{\text d}\omega\left[\frac{a_{ii}+b_{ii}k_2^2+c_{ii}k_2^4}{(|k_1|^2+k_2^2)^2}
\right]\frac{\cos(2k_2 R)}{k_2R},
\label{eq:simplechi}
\end{align}
with $\chi_0=2m_0\pi^2/(\hbar^2\gamma_1d^2)$ and coefficients
\begin{align}
a_{xx}&=a_{yy}=Z\left[12\bar\gamma(\omega-\omega_0)-5(\omega-\omega_0)^2\right],\nonumber\\[2mm]
b_{xx,yy}&=Z\left\{\left[10+4\bar\gamma(1\pm 3\alpha)\right](\omega-\omega_0)-12\bar\gamma(1+\bar\gamma)\right\},\nonumber\\[2mm]
c_{xx,yy}&=Z\left\{\bar{\gamma } \left[\bar{\gamma}-9 \alpha ^2 \bar{\gamma }-4\mp12 \alpha  \left(\bar{\gamma }+1\right) 
\right]-5 \right\}, \nonumber \\[2mm]
a_{zz}&=Z\left[36\bar\gamma(\omega-\omega_0)-5(\omega-\omega_0)^2-72\bar\gamma^2\right],\nonumber\\[2mm]
b_{zz}&=Z\left[2(5-4\bar\gamma)(\omega-\omega_0)+36\bar\gamma(\bar\gamma-1)\right],\nonumber\\[2mm]
c_{zz}&=Z\left\{\bar{\gamma } \left[\bar{\gamma}\left(9 \alpha ^2-5\right)+8\right]-5\right\},  
\end{align}
%
where $Z=\frac{1}{(8\pi^2)}\left[\left(3 \alpha ^2+1\right) \bar{\gamma}^2-1\right]^{-2}$ and $k_{1,2}$ are given in Eq.~(\ref{eq:Poles}) with $\bar{\omega}\to\omega$.
We note that in the limit of zero HH-LH mixing, $\alpha\to 0$, we find that the elements $\chi_{xx}({\bf R})=\chi_{yy}({\bf R})$ decay exponentially.
This can be understood from the form of the Green's function in Eq.~(\ref{eq:Geensfunreal}) and the spin susceptibility tensor 
in Eqs.~(\ref{eq:Trxx})-(\ref{eq:Trzz}).
As there are only HH states mediating the RKKY interaction but no mixing with LH states, only the first term
in Eq.~(\ref{eq:Trzz}) can be non-vanishing, and every single term in the sums vanishes identically (after integration).
The spin susceptibility tensor element $\chi_{zz}({\bf R})$, on the other hand, is non-vanishing and coincides with the RKKY range function
of a 2DEG in the limit $\alpha\to 0$.

Furthermore, we note that due to the axial symmetry of the Hamiltonian, the result for the
spin susceptibility tensor for arbitrary $\phi_R$ is obtain by an orthogonality transformation of 
$\chi_{ii}(R,\phi_R=0)$ with a rotation about the $z$-axis with an angle $\phi_R$.   
Of course, this leaves $\chi_{zz}(\bf{R})$ invariant, and one simply has to 
transform the coefficient matrices in Eq.~(\ref{eq:simplechi}), e.g., $a_{ii}\to O\cdot\text{diag}(a_{ii})\cdot O^T$ etc. 

Moreover, we find that the largest in-plane components of spin suscepetibility tensor
are obtained in the case where the axis connecting two localized impurities coincide with their spin-quantization axis, i.e., 
$|\chi_{xx}|$ is largest [smallest] for $\phi_R=0$ [$\phi_R=\pi/2$], whereas for the magnitude of $\chi_{yy}$ the opposite relation holds.

\section{Numerical Results\label{sec:numerics}}

Now we turn to a numerical analysis to study the dependence of the spin susceptibility $\chi_{ij}({\bf R})$, Eq.~(\ref{eq:Matsub}), on the 
Fermi energy $E_{F}$ and the band structure parameters $\bar\gamma$ and $\alpha$.
In accordance with the calculation of the spin susceptibility in the previous section, we
consider only cases where only the lowest HH-like subband is occupied. 

We have checked that the result based on the analytical approach given in the present paper agrees with the numerical
method of calculating the spin susceptibility by means of eigenbasis functions of the Hamiltonian, i.e., by employing
the Lehmann representation for the Green's functions\cite{hole2DLind}.
\subsection{Spin susceptibility of GaAs}

\begin{figure}[t]
\includegraphics[width=7.cm]{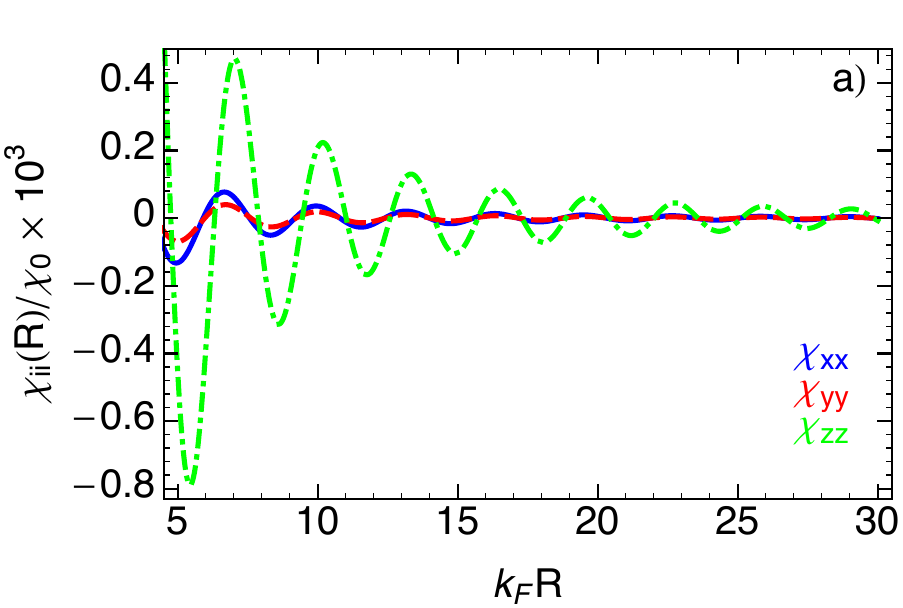}
 \includegraphics[width=7.cm]{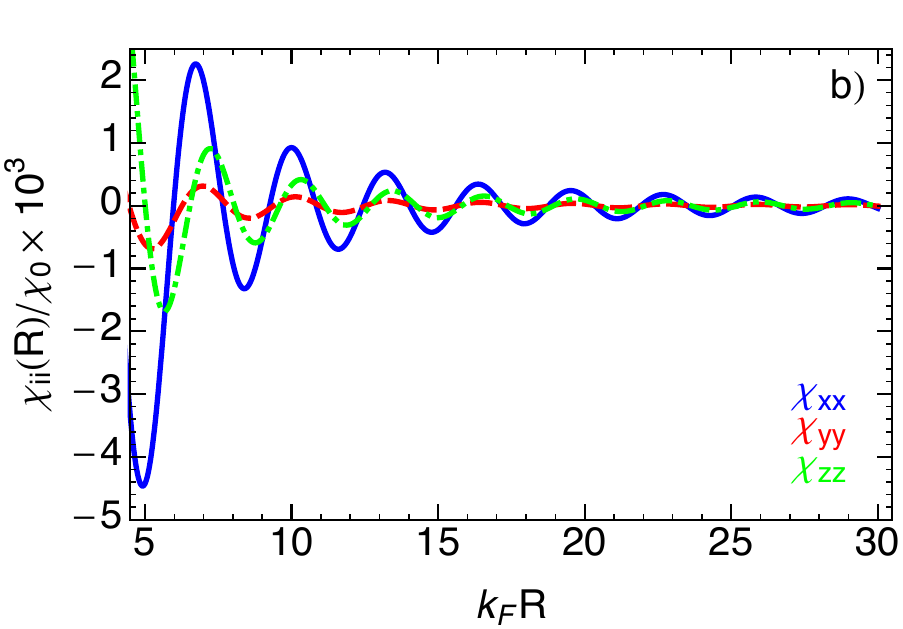}
  \caption{The spin susceptibility tensor entries $\chi_{ij}({\bf R})$ as a function of $k_FR$ (with $\phi_R=0$), for
  (a) $\varepsilon_F=0.9$, and (b) $\varepsilon_F=1.4$.}                                                 
  \label{fig:FriedelOsc}
\end{figure}

We start by presenting results of the RKKY range function for the case of a [001]-grown GaAs heterostructure where 
the corresponding band structure parameter values are $\bar\gamma=0.31$ and
$\alpha=1.2$. In the following examples we vary the Fermi density $E_F=E_0 \varepsilon_{F}$ with 
the dimensionless parameters  $\varepsilon_{F}$.
In Fig.~\ref{fig:FriedelOsc}, we show $\chi_{ij}({\bf R})$ (with $\phi_R=0$) as function of $k_FR$, for the Fermi energies 
$\varepsilon_{F}=0.9$ and $\varepsilon_{F}=1.4$, respectively.
In both plots, the Friedel oscillations decay as $R^{-2}$, which is the usual result for two-dimensional systems.
By comparing Figs.~\ref{fig:FriedelOsc}(a) and (b), we see that for lower hole densities $\chi_{zz}(R)$ dominates, whereas for increased
hole density $\chi_{xx}(R)$ becomes the dominant entry of the spin susceptibility tensor.
In Fig.~\ref{fig:FriedelOsc}(b), we can also clearly see the strong in influence of the HH-LH mixing parameter $\alpha$ which 
gives rise to $\chi_{xx}(R)\gg \chi_{yy}(R)$.

\subsection{Full parameter dependence of spin susceptibility}

Now we will consider scenarios where in addition to the Fermi energy the band structure parameters $\bar{\gamma}$ and $\alpha$ are varied.
Obviously most values will not correspond to actual semiconductor materials.
As we will see however, such an approach allows us to elucidate the influence of HH-LH mixing on the spin susceptibility tensor entries.
Again we discuss density ranges where only the lowest HH-like subband is occupied.
The next-to-lowest subband is either the lowest LH-like subband or the next-to-lowest HH-like subband.
Which of the two situation is realized depends on the value of the HH-LH splitting parameter $\bar{\gamma}$,
which is related to the corresponding band edge energies by $\varepsilon=1+2\bar\gamma$  and $\varepsilon=4(1-2\bar\gamma)$,
respectively. Thus we impose the following constraint on the Fermi energy:
$\varepsilon_{F}<{\rm min}\{1+2\bar\gamma,4(1-2\bar\gamma)\}$. 

\begin{figure}[t]
\begin{center}
\includegraphics[width=6.cm]{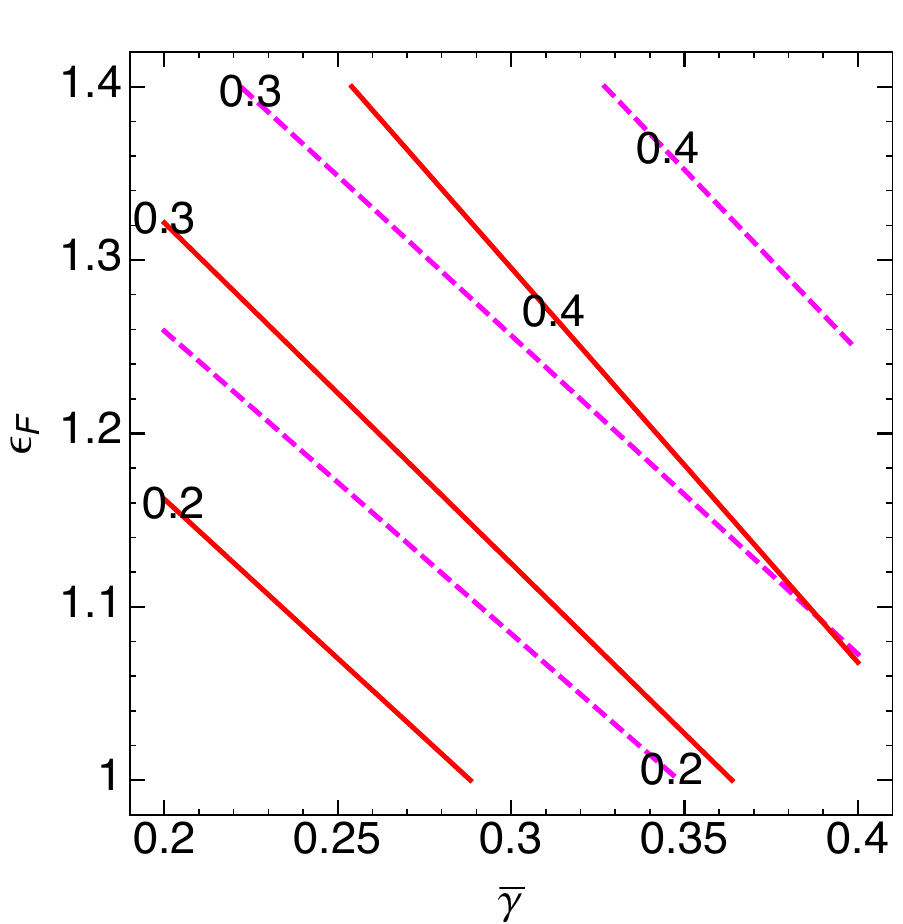}
\caption{Contours of $\sin^2\theta_{HL}$, Eq.~(\ref{eq:mixing}), in the $\bar\gamma$-$\varepsilon_{F}$ plane, for $\alpha=1.0$ (dashed lines) and $\alpha=1.2$ (solid lines).}                                                 
\label{fig:SinTheta}
\end{center}  
\end{figure}

To study the influence of HH-LH mixing for this general case, it is convenient to define the following 
HH-LH mixing angle\cite{hole2DLind} 
\begin{align}
\sin\theta_{HL}=   \frac{\sqrt{3} \alpha  k_{F}^2}{\sqrt{3 \alpha ^2
   k_{F}^4+\left(\sqrt{3 \alpha ^2
   k_{F}^4+\left(k_{F}^2-2\right)^2}-k_{F}^2+2\right)^2}}~,
   \label{eq:mixing}
\end{align}
which depends only on $\alpha$ and $k_{F}$. 
The Fermi wave vector $k_{F}$ depends in turn on the Fermi energy $\varepsilon_{F}$ and the
band structure parameters $\bar\gamma$ and $\alpha$, see Eq.~(\ref{eq:Poles}).
The modulus squared of $\sin\theta_{HL}$ tells us the amount of light hole character of the lowest HH-like band and is therefore
a measure for HH-LH mixing.
In order to show the $\sin^2\theta_{HL}$ dependence on the Fermi energy as well as on the parameters 
$\bar\gamma$ and $\alpha$, we plot in Fig.~\ref{fig:SinTheta} $\sin^2\theta_{HL}$ as a function of $\bar\gamma$  and $\varepsilon_{F}$
for $\alpha=1.0$ (dashed lines) and $\alpha=1.2$ (solid lines).
It can be seen that $\sin^2\theta_{HL}$ is monotonically increasing with $\varepsilon_{F}$, $\bar\gamma$, and $\alpha$.

\subsubsection{Anisotropy between $\chi_{xx}$ and $\chi_{yy}$}

\begin{figure}[t]
\includegraphics[width=7.cm]{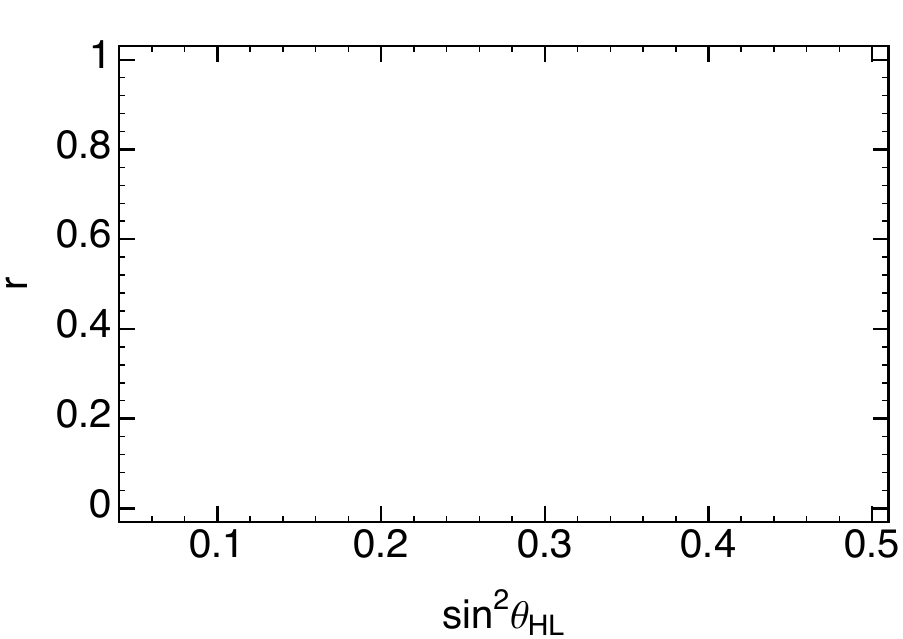}
  \caption{The ratio $r$, Eq.~(\ref{eq:ratioaniso}), versus $\sin^2\theta_{HL}$.}                                                 
  \label{fig:Ani}
\end{figure}

Now we analyze the size of in-plane anisotropy due to HH-LH mixing, where we define the following ratio
to quantify the deviation from the isotropic case:
\begin{align}
r\equiv  \frac{\chi_{xx}({\bf R})-\chi_{yy}({\bf R})}{\chi_{xx}({\bf R})+\chi_{yy}({\bf R})}~.
\label{eq:ratioaniso}
\end{align}
We then randomly generate 1000 number triples of the structure parameters within their respective ranges $\bar\gamma\in(0.2,0.4)$ and $\alpha\in (0.9,1.4)$ and the Fermi energy in the range $\varepsilon_F\in (0.9,1.5)$.
Using these number triples, we calculate $r$ and $\sin\theta_{HL}$. The result is displayed 
in Fig.~\ref{fig:Ani}, where we plot $r$ (for $k_FR=10$ and $\phi_R=0$) versus $\sin^2\theta_{HL}$.
We find a clear correlation between the in-plane anisotropy and the HH-LH mixing angle, showing that $r$ is a monotonically increasing function
of $\sin^2\theta_{HL}$. The in-plane anisotropy can go up to 90\%.
We note that this result is not very sensitive to the choice for $k_F R$, provided we take values in the vicinity of a maxima of the Friedel oscillations (and $k_F R\gg1$).

\subsubsection{Anisotropy between $\chi_{xx}$ and $\chi_{zz}$}

\begin{figure}[t]
 \includegraphics[width=6.cm]{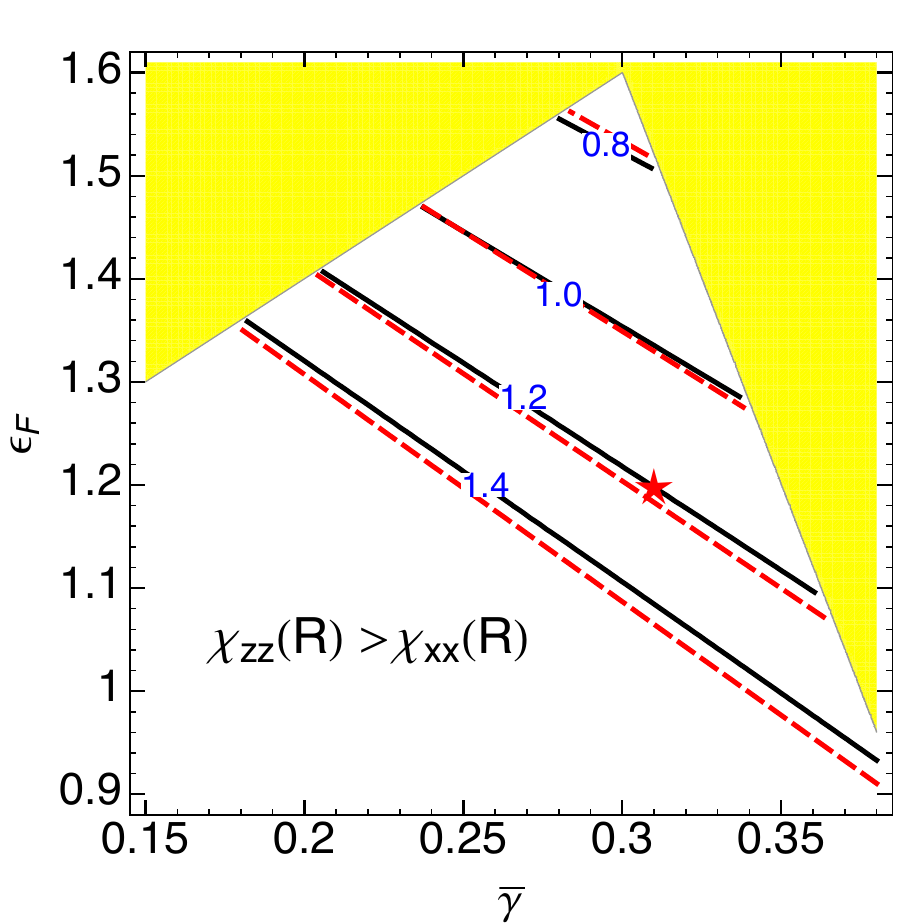}
  \caption{Boundaries in the $\bar\gamma$-$\varepsilon_F$ plane where $\chi_{xx}({\bf{R}})=\chi_{zz}({\bf{R}})$, for
  HH-LH mixing parameter values $\alpha=0.8,1.0,1.2,1.4$ (black lines). For the area above [below] the respective line we 
  have $\chi_{xx}({\bf{R}})>\chi_{zz}({\bf{R}})$ [$\chi_{xx}({\bf{R}})<\chi_{zz}({\bf{R}})$].
  Corresponding to each value of $\alpha$ the red dashed lines are the contour lines for $\sin^2\theta_{HL}=0.35$.
  The yellow area is excluded by the condition that only the lowest subband is occupied.
  The red star symbol indicates the point for GaAs, where for all $\varepsilon_F\gsim 1.2$ we have $\chi_{xx}({\bf{R}})>\chi_{zz}({\bf{R}})$.}                                                 
  \label{fig:Contour1}
\end{figure}

\begin{figure}[t]
 \includegraphics[width=6.cm]{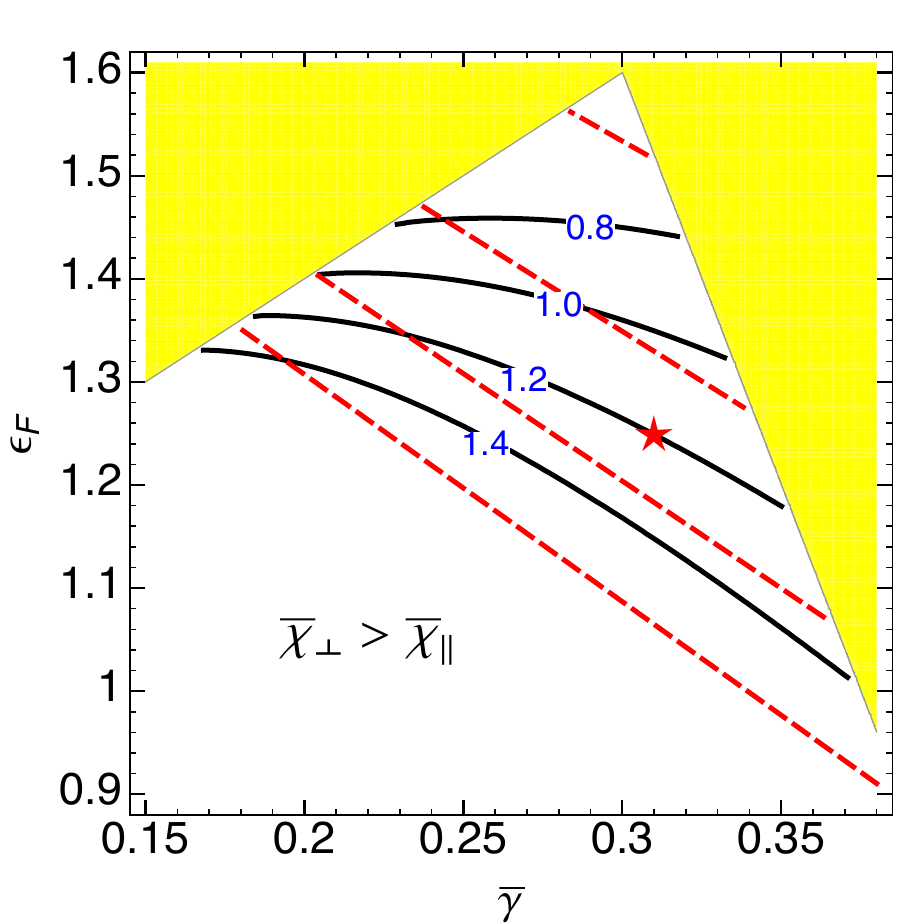}
  \caption{Same as in Fig.~\ref{fig:Contour1} but for the integrated spin susceptibility elements $\overline{\chi}_{\perp}$ and $\overline{\chi}_{\parallel}$, Eq.~(\ref{eq:IntegralR}).}                                                 
  \label{fig:Contour2}
\end{figure}

We have seen in Fig.~\ref{fig:FriedelOsc} that the dominance between $\chi_{xx}({\bf R})$ and $\chi_{zz}({\bf R})$ changes with the value of the Fermi energy.
This can be attributed to HH-LH mixing because by increasing the Fermi energy (density), hole states with larger wave vectors are populated which exhibit 
a stronger HH-LH mixing.\cite{RolandBook} However, for Fig.~\ref{fig:FriedelOsc} the band structure parameters $\bar{\gamma}$ and $\alpha$ were held fixed 
and it is not clear if other values would give rise to a different behaviour. 
To answer this question whether HH-LH mixing is indeed the underlying mechanism for the change from easy-axis to easy-plane dominance,
we show in Fig.~\ref{fig:Contour1} the boundary lines of $\chi_{xx}({\bf R})=\chi_{zz}({\bf R})$ in the $\bar\gamma$-$\varepsilon_{F}$ plane for various
values of $\alpha$, choosing $k_{\text F}R=10$ and $\phi_R=0$. 
Below the lines we have $\chi_{zz}({\bf R})>\chi_{xx}({\bf R})$ and above the lines the opposite relation.
The dashed lines are contour lines for $\sin^2\theta_{HL}=0.35$ associated to each value of $\alpha$. 
As can be seen from Fig.~\ref{fig:Contour1}, the boundary lines where the transition $\chi_{xx}({\bf R})<\chi_{zz}({\bf R})$ to
$\chi_{xx}({\bf R})>\chi_{zz}({\bf R})$ occurs
almost coincides with the corresponding contour of $\sin^2\theta_{HL}= 0.35$. 
Fig.~\ref{fig:Contour1} implies that an increase of the amount of HH-LH mixing also entails an increase 
of $\chi_{xx}({\bf R})/\chi_{zz}({\bf R})$ and determines the easy-axis versus easy-plane dominance of the impurity spins.
Moreover, Fig.~\ref{fig:Contour1} shows that there is an approximately universal
value for the HH-LH mixing angle $\sin^2\theta_{HL}\sim 0.35$ at which the phase transition occurs.
The behaviour of the easy-axis versus easy-plane components of the spin susceptibility tensor
can be understood intuitively by considering the influence of a in-plane magnetic field on a two-dimensional hole gas.
An in-plane magnetic field has a suppressed coupling to HH states\cite{Winkler,RolandBook}, which in turn implies a tiny Zeeman-splitting. 
On the contrary, the coupling of an in-plane magnetic field to LH states is not suppressed.
Thus, these features of HH and LH states get interchanged when HH-LH mixing is promoted, and
clearly leaves an imprint in the spin susceptibility tensor.
Consequently, one could conjecture that the easy-plane components $\chi_{xx}$ and $\chi_{yy}$ are increased with respect to 
the easy-axis component $\chi_{zz}$ when HH-LH mixing increases. 
It is however worth emphasizing that the transition happens not for $\sin^2\theta_{HL}\sim 0.5$, 
as one would naively expect from this argument, but for a much lower value.

So far we have considered the case of two isolated impurities and their exchange interaction mediated by the
spin susceptibility tensor as given in Eqs.~(\ref{eq:Trxx})-(\ref{eq:Trzz}).
In semiconductor systems with a high density of magnetic impurities it is useful to average over the distances of all impurities
assuming that they are randomly but on the average homogeneous distributed.
This corresponds to taking the mean field limit in the calculation of the Curie temperature\cite{Dietl}.
In such a way the discrete sum can be replaced by an integral\cite{footnote}
\begin{equation}
\overline{\chi}_{ii}=n^{\text{imp}}\int ~{\text d}{\bf R}~ {\chi_{ii}}({\bf R})~,
\label{eq:IntegralR}
\end{equation}
where $n^{\text{imp}}$ denotes the density of impurities and we define 
$\overline{\chi}_{zz}\equiv\overline{\chi}_\perp$ and $\overline{\chi}_{xx}=\overline{\chi}_{yy}\equiv\overline{\chi}_\parallel$ 
since the angular part of the in-plane components drops out after the integration over $\phi_R$, see Eqs.~(\ref{eq:Trxx}) and (\ref{eq:Tryy}).
In Fig.~\ref{fig:Contour2}, we show the boundary lines of $\overline{\chi}_{\parallel}=\overline{\chi}_{\perp}$ in the $\bar\gamma$-$\varepsilon_{F}$ plane for the same
values of $\alpha$ as in Fig.~\ref{fig:Contour1}. 
Again we include the contours $\sin^2\theta_{HL}=0.35$ as in Fig.~\ref{fig:Contour1}. 
In comparison with Fig.~\ref{fig:Contour1}, we see that an averaging over the distance 
leads to a distortion of the linear character of the boundary lines.
This is mainly due to short distance contributions which show a different behaviour.
As a result, the boundary lines do not follow the linear behaviour of $\sin^2\theta_{LH}$ over the whole parameter region shown.
However, we still find that $\overline{\chi}_{\parallel}/\overline{\chi}_{\perp}$ is monotonically increasing with the Fermi energy $\varepsilon_F$ and the 
structure parameters $\alpha$ and $\bar{\gamma}$, and thus is correlated with the HH-LH mixing angle.
Exceptions to this behaviour are found for $0.6\lsim\alpha\lsim0.8$, where an increase of $\bar\gamma$ does not
yield larger $\overline{\chi}_{\parallel}/\overline{\chi}_{\perp}$.
Clearly, also the universal behaviour of Fig.~\ref{fig:Contour1} is lost and a transition from easy-axis to easy-plane dominance does not happen globally close to a particular value of $\sin^2\theta_{HL}$.
For example, in the cases $\alpha=1.4$ and $\bar{\gamma}\gsim 0.3$, the phase transition occurs at $\sin^2\theta_{HL}\sim 0.39$.
Whereas for $\alpha=0.8$, the value for $\sin^2\theta_{HL}$ can be around 0.32 to still obtain $\overline{\chi}_{\parallel}/\overline{\chi}_{\perp}>1$.

\section{Summary \label{sec:sum}}

We have calculated and analyzed the spin susceptibility tensor of a homogeneous 2D hole gas, based on the Luttinger-model description of the
lowest valence band within axial approximation. In such a way analytical results can be obtained that comprise the explicit dependence on 
the relevant band structure parameters. Our formulae show the important influence of HH-LH mixing on the elements of the spin susceptibility tensor.
We find strong anisotropies both among the easy-plane components as well as among the easy-plane and easy-axis components.
Moreover, we have pointed out
that these anisotropies are intimately connected to the HH-LH mixed character of the hole states. 
In particular, we find that the anisotropy between easy-plane components depends only on the amount of LH character
in the lowest HH-like band, characterized by $\sin^2\theta_{HL}$, Eq.~(\ref{eq:mixing}).
Also, we find an almost universal value for $\sin^2\theta_{HL}$ for the switching from easy-axis to easy-plane aligned
impurity spins.
In contrast, we recover the well-known result of an 2DEG in the limit of zero HH-LH mixing, with impurity spins aligned perpendicular to the
quantum-well.

\acknowledgements

The author is indebted to M. Governale, R. Winkler and U. Z\"ulicke for numerous helpful and interesting discussions.

%


\begin{thebibliography}{99}
\bibitem{footnote}
The positional averaging done here serves as a qualitative estimate of the
 the easy-axis versus easy-plane behaviour.
It should be however noted that in order to arrive at a more quantitative estimate of this feature,
subtleties in doing the sum would need to be taken into account, see for instance Refs.~\onlinecite{Brey,Fiete}.
\end{thebibliography}
\end{document}